\documentclass[prd,superscriptaddress,amsfonts,amssymb,amsmath,showpacs,onecolumn]{revtex4-2}
\usepackage{bm}
\usepackage{amsfonts}
\usepackage{latexsym}
\usepackage{graphicx}
\usepackage{amsmath}
\usepackage{palatino}
\usepackage{mathpazo}
\usepackage{textcomp}
\usepackage{float}
\usepackage{booktabs}
\usepackage{dcolumn}
\usepackage{booktabs}
\usepackage{multirow}
\usepackage{hyperref}
\hypersetup{colorlinks,citecolor=blue}
\usepackage{amsmath}
\usepackage{xcolor}
\usepackage{orcidlink}
\usepackage{epsfig}
\usepackage{caption}
\usepackage{subcaption}
\usepackage{commath}
\def\jnl@style{\it}
\def\aaref@jnl#1{{\jnl@style#1}}

\def\aaref@jnl#1{{\jnl@style#1}}

\def\aj{\aaref@jnl{AJ}}                   % Astronomical Journal
\def\apj{\aaref@jnl{ApJ}}                 % Astrophysical Journal
\def\apjl{\aaref@jnl{ApJ}}                % Astrophysical Journal, Letters
\def\apjs{\aaref@jnl{ApJS}}               % Astrophysical Journal, Supplement
\def\apss{\aaref@jnl{Ap\&SS}}             % Astrophysics and Space Science
\def\aap{\aaref@jnl{A\&A}}                % Astronomy and Astrophysics
\def\aapr{\aaref@jnl{A\&A~Rev.}}          % Astronomy and Astrophysics Reviews
\def\aaps{\aaref@jnl{A\&AS}}              % Astronomy and Astrophysics, Supplement
\def\mnras{\aaref@jnl{Mon.~Not.~Roy.~Astron.~Soc.}}             % Monthly Notices of the RAS
\def\prd{\aaref@jnl{Phys.~Rev.~D}}        % Physical Review D
\def\prc{\aaref@jnl{Phys.~Rev.~C}}  % Physical Review C
\def\prl{\aaref@jnl{Phys.~Rev.~Lett.}}    % Physical Review Letters
\def\qjras{\aaref@jnl{QJRAS}}             % Quarterly Journal of the RAS
\def\skytel{\aaref@jnl{S\&T}}             % Sky and Telescope
\def\ssr{\aaref@jnl{Space~Sci.~Rev.}}     % Space Science Reviews
\def\zap{\aaref@jnl{ZAp}}                 % Zeitschrift fuer Astrophysik
\def\nat{\aaref@jnl{Nature}}              % Nature
\def\aplett{\aaref@jnl{Astrophys.~Lett.}} % Astrophysics Letters
\def\apspr{\aaref@jnl{Astrophys.~Space~Phys.~Res.}} % Astrophysics Space Physics Research
\def\physrep{\aaref@jnl{Phys.~Rep.}}      % Physics Reports
\def\physscr{\aaref@jnl{Phys.~Scr}}       % Physica Scripta
\def\commat{\aaref@jnl{Comm.~Math.~Phys.}}              % Communications in Mathematical Physics
\def\science{\aaref@jnl{Science}}               % Science
\def\cqg{\aaref@jnl{Classical Quant.~Grav.}}            % Classical and Quantum Gravity
\def\jpcs{\aaref@jnl{JPCS}}                                     % Journal of Physics Conference Series
\def\ijmpd{\aaref@jnl{Int.~J.~Mod.~Phys.~D}}                    % International Journal of Modern Physics D
\def\grg{\aaref@jnl{Gen.~Relat.~Gravit.}}               % General Relativity and Gravitation
\def\rpp{\aaref@jnl{Rep.~Prog.~Phys.}}          % Reports on Progress in Physics
\def\npa{\aaref@jnl{Nucl.~Phys.~A}}        % Nuclear Physics A
\def\lrr{\aaref@jnl{Living Rev.~Rel.}}                   % Living reviews in relativity
\def\jcap{\aaref@jnl{J.~Cosmology Astropart.~Phys.}}    % Journal of cosmology and astroparticle physics
\def\rmp{\aaref@jnl{Rev.~Mod.~Phys.}}   %Reviews of modern physics
\def\epjc{\aaref@jnl{Eur.~Phys.~J.~C}}

\allowdisplaybreaks[1]
\begin{document}

\color{black}       %% For one column

\title{Non-singular bouncing cosmology in $f(T, \mathcal{T})$ gravity with energy condition violations}

\author{A. Zhadyranova\orcidlink{0000-0003-1153-3438}} 
\email[Email: ]{a.a.zhadyranova@gmail.com}
\affiliation{General and Theoretical Physics Department, L. N. Gumilyov Eurasian National University, Astana 010008, Kazakhstan.}

\author{M. Koussour\orcidlink{0000-0002-4188-0572}}
\email[Email: ]{pr.mouhssine@gmail.com}
\affiliation{Department of Physics, University of Hassan II, Casablanca, Morocco.}

\author{V. Zhumabekova\orcidlink{0000-0002-7223-5373}}
\email[Email: ]{zh.venera@mail.ru}
\affiliation{Theoretical and Nuclear Physics Department, Al-Farabi Kazakh National University, Almaty, 050040, Kazakhstan.}

\author{N. Zhusupova\orcidlink{0009-0001-8900-0829}}
\email[Email: ]{zhnaz88@gmail.com}
\affiliation{Department of Physics, Abai Kazakh National Pedagogical University, Almaty, 050010, Kazakhstan.}

\author{S. Muminov\orcidlink{0000-0003-2471-4836}}
\email[Email: ]{sokhibjan.muminov@gmail.com}
\affiliation{Mamun University, Bolkhovuz Street 2, Khiva 220900, Uzbekistan.}

\author{J. Rayimbaev\orcidlink{0000-0001-9293-1838}}
\email[Email: ]{javlon@astrin.uz}
\affiliation{University of Tashkent for Applied Sciences, Gavhar Str. 1, Tashkent 100149, Uzbekistan.}
\affiliation{National University of Uzbekistan, Tashkent 100174, Uzbekistan.}

%%%%%%%%%%%%%%%%%%%%%%%%%%%%%%%%%%%%  DATE  %%%%%%%%%%%%%%%%%%%%%%%%%%%%%%%%%%%%
%\date{\today}

\begin{abstract}

The singularity and inflationary problems have posed significant challenges for understanding the universe's origin and evolution. Bouncing cosmology has emerged as a promising alternative to standard cosmological models, offering a non-singular approach to early universe dynamics by facilitating a "bounce" rather than a singular beginning. In this study, we explore the feasibility of modeling specific bouncing scenarios within the framework of $ f(T, \mathcal{T}) $ gravity, allowing for a comprehensive coupling between the torsion scalar $T$ and the trace of the energy-momentum tensor $\mathcal{T}$. We analyze two $f(T, \mathcal{T})$ models: a linear model $f(T, \mathcal{T}) = \alpha T + \beta \mathcal{T}$ and a non-linear model $f(T, \mathcal{T}) = \alpha \sqrt{-T} + \beta \mathcal{T}$, with a parameterized scale factor $a(t) = \sqrt{a_0^2 + \gamma^2 t^2}$ to capture the bounce behavior. The analysis confirms a cosmic bounce at $t = 0 $, where the Hubble parameter $H = 0$ signals a transition from contraction to expansion. A crucial condition for achieving the bounce is the violation of the null energy condition (NEC) near the bounce, enabling the equation of state (EoS) parameter to enter the phantom region ($\omega < -1$). Both models exhibit an increase in energy density as the universe approaches the bounce, peaking at the bounce epoch and then decreasing post-bounce. Pressure remains negative throughout, with the EoS parameter crossing into the phantom region near the bounce in both positive and negative time zones. Our findings show that NEC and strong energy condition (SEC) violations are essential for the non-singular bounce, while the dominant energy condition (DEC) is satisfied, ensuring a consistent matter distribution. These results indicate that both linear and non-linear $f(T, \mathcal{T})$ models effectively replicate the critical features of a bouncing cosmology, offering valuable approaches for addressing the singularity and inflationary challenges in cosmology.

\end{abstract}

\maketitle

\section{Introduction}\label{sec1}

Recent astrophysical observations, such as those from Type Ia supernovae (SNe-Ia) \cite{R1, R2}, baryon acoustic oscillations (BAO) \cite{R3, R4}, the wilkinson microwave anisotropy probe (WMAP) \cite{R5}, and cosmic microwave background (CMB) radiation \cite{R6, R7}, have shown that the expansion of the universe is currently accelerating. These observations suggest that the universe is flat, homogeneous, isotropic, and can be described by a Friedmann-Lemaître-Robertson-Walker (FLRW) spacetime. The discovery of cosmic acceleration has spurred the scientific community to move beyond Einstein’s general relativity (GR) to explore other mechanisms that could explain this behavior. In the framework of GR, the cosmological constant $\Lambda$ is introduced to account for the presence of dark energy (DE), a mysterious form of energy responsible for the accelerated expansion. However, this model faces several challenges, such as the fine-tuning problem, the coincidence problem, and the fact that DE is only observed at cosmological scales rather than at Planck scales \cite{R8,R9}. These issues have led to the development of alternative modified theories of gravity, which attempt to address these shortcomings and provide a more comprehensive understanding of cosmic phenomena.

Currently, two main approaches are widely used to explain the universe's accelerated expansion. The first is the introduction of DE with negative pressure into Einstein’s field equations, which has led to the proposal of various models, such as phantom, quintessence, and k-essence \cite{R10,R11,R12}. The second approach modifies the gravitational theory itself, particularly at large scales, to account for the observed cosmic acceleration. Modified gravity theories have shown promising results in explaining both early-time inflationary behavior and late-time acceleration \cite{R20}. Among these modified theories, $f(R)$ gravity and $f(T)$ gravity have gained significant attention. In $f(R)$ gravity, the curvature scalar $R$ in the Einstein-Hilbert action is replaced by an arbitrary function $ f(R)$, and the theory is described using the Levi-Civita connection, which characterizes a spacetime with zero torsion and non-metricity but non-vanishing curvature \cite{R21,R22,R23}. On the other hand, $f(R, \mathcal{T})$ gravity generalizes $f(R)$ gravity by coupling the Ricci scalar $R$ with the trace of the energy-momentum tensor $\mathcal{T}$ \cite{RR24,RR25,RR26,RR27}. Similarly, $f(R, G)$ gravity extends the theory by including the Gauss-Bonnet term $G$ in the action \cite{RR28,RR29}. Recent developments in cosmology have highlighted limitations in describing gravitational interactions on cosmological scales with traditional Riemannian geometry, which performs well at smaller scales, such as within the solar system. This has led to the exploration of alternative geometric frameworks, including Weitzenb\"{o}ck spaces, introduced by Weitzenb\"{o}ck himself to extend the foundations of geometry for broader applications in gravitational theories \cite{RR30}. A Weitzenb\"{o}ck manifold is characterized by a covariantly constant metric tensor, $ g_{\sigma\lambda}$, a non-zero torsion tensor $ T^\mu_{\sigma\lambda}$, and a vanishing Riemann curvature tensor $R^\mu_{\nu\sigma\lambda}$. This combination yields a geometry in which the manifold reduces to Euclidean space in the absence of torsion, while torsion itself varies spatially across the manifold to enable alternative representations of gravitational effects. One unique feature of Weitzenb\"{o}ck geometry is its zero-curvature property, which supports the concept of distant parallelism, also known as teleparallelism. This idea attracted Einstein’s attention as a foundation for a unified theory of electromagnetism and gravity, leading to the development of teleparallel gravity, which describes gravity through torsion rather than curvature \cite{RR31}. In teleparallel gravity, gravitational effects are captured using a tetrad field $ e_i^{\mu}$ instead of the traditional spacetime metric $ g_{\mu\nu}$. The tetrads generate torsion, allowing gravity to be fully described in terms of torsion alone. This approach gave rise to the teleparallel equivalent of GR (TEGR) and eventually to $f(T)$ gravity, where torsion replaces curvature to yield a flat spacetime framework \cite{RR32,RR33,RR34}. A significant advantage of $f(T)$ gravity lies in its second-order field equations, offering a simpler formulation compared to the fourth-order equations in $f(R)$ gravity. This theory has proven effective in explaining late-time cosmic acceleration without resorting to DE, making it a compelling alternative for addressing open questions in cosmology. The application of $f(T)$ gravity to cosmological and astrophysical phenomena has therefore gained considerable interest, as it provides a promising foundation for models of cosmic evolution and the dynamics of large-scale structures in the universe \cite{RR35,RR36,RR37,RR38,RR39,RR40,RR41}. The $f(T, \mathcal{T}) $ gravity model, introduced by Harko et al. \cite{Harko_fTT}, has garnered interest for its dual focus on the torsion scalar $T$ and the energy-momentum trace $\mathcal{T}$. This theory builds upon teleparallel gravity, allowing for a richer description of gravitational dynamics by incorporating interactions between torsion and the energy-momentum contributions from matter fields. The $f(T, \mathcal{T})$ model has been applied in various cosmological and astrophysical contexts, where its implications for cosmic expansion, structure formation, and compact objects have been actively explored. The cosmological relevance of $ f(T, \mathcal{T})$ gravity has been examined extensively. Junior et al. \cite{TT17} analyzed the thermodynamics, stability, and reconstruction of the $\Lambda$CDM model within this framework, showing that it aligns with observational data while perhaps providing information about how classical and quantum gravity behaves. Furthermore, Harko et al. \cite{HARK} extended $f(T)$ gravity by introducing a non-minimal coupling between torsion and matter, enhancing the model's ability to capture the intricate interactions in cosmological dynamics. Similarly, Momeni and Myrzakulov \cite{TT18} explored the cosmological reconstruction of $f(T, \mathcal{T})$, shedding light on its potential to describe diverse evolutionary scenarios in the universe. Studies of structure formation in $f(T, \mathcal{T})$ gravity have further deepened our understanding of cosmic evolution. Farrugia and Said \cite{TT19} analyzed the growth factor of cosmic structures in this framework, while Pace and Said \cite{TT20} employed a perturbative approach to investigate the behavior of compact objects like neutron stars. These studies collectively emphasize the potential of $f(T, \mathcal{T})$ gravity to provide a unified framework for explaining cosmic expansion, structure formation, and compact object dynamics, marking it as a promising candidate for addressing key questions in modern cosmology.

The inflationary model suggests that the universe experienced a rapid expansion in its earliest moments, smoothing out initial irregularities and setting the stage for its observed large-scale structure. However, this model does not fully account for the origin of the universe, as it assumes a pre-existing singularity prior to inflation. To address this limitation, the matter bounce scenario has emerged as a compelling alternative, proposing that the universe initially underwent a contraction phase before experiencing a bounce, which then led to the current expansion phase \cite{MB-1,MB-2,Haro}. This approach allows for a Universe that avoids a singularity, replacing it with a bounce that generates causal fluctuations, potentially seeding structure formation. The matter bounce scenario introduces an initial matter-dominated contraction phase, which eventually reverses into expansion through a non-singular bounce, challenging the need for a singular beginning. This non-singular cosmological model, however, often requires a violation of the null energy condition (NEC), which has been demonstrated in certain modified gravity theories, such as generalized Galileon models \cite{MB-3}. The concept of a big bounce replacing the Big Bang singularity has drawn substantial interest in the field of modified gravity \cite{Zuba,Bhat,Yous,K_B1,K_B2,K_B3}, inspiring new ways to model the early universe and the transition from contraction to expansion. Numerous studies have explored the implications of bouncing cosmologies within various modified gravity frameworks, including $f(T)$ gravity. For instance, Cai et al. \cite{B_T1} investigated a matter bounce cosmology within the framework of $f(T)$ gravity, while Rodrigues and Junior \cite{B_T2} examined black-bounce solutions within $f(T)$ gravity, analyzing how these configurations can address singularities in black hole models. Further work by Amorós et al. \cite{B_T3} explored matter bounce scenarios in loop quantum cosmology derived from $f(T)$ gravity, and Skugoreva and Toporensky \cite{B_T4} studied bouncing solutions using the power-law $f(T)$ model. Other studies have looked into bouncing models addressing potential future singularities, collectively underscoring the versatility of modified gravity in addressing early-universe phenomena and advancing our understanding of the Universe's origin and evolution. Despite the successes of inflationary models, they are not without challenges, namely, the trans-Planckian problem \cite{Martin/2001}, fine-tuning of initial conditions \cite{R9}, and difficulty in embedding in quantum gravity frameworks. Bouncing cosmologies provide an alternative paradigm where the universe undergoes a contraction phase followed by a regular bounce into expansion, avoiding the initial singularity. In this context, $f(T,\mathcal{T})$ gravity offers a compelling framework: it allows for NEC violation through matter-geometry coupling without introducing ghost instabilities typical in higher-order curvature theories. Furthermore, since torsion-based theories are dynamically second-order, they are more tractable for bounce analysis compared to $f(R,\mathcal{T})$ \cite{RR24} or loop quantum cosmology (LQC) scenarios \cite{Bojowald/2001}. In this work, we explore the possibility of a non-singular bouncing universe within the $ f(T, \mathcal{T})$ gravity framework. The structure of this paper is as follows: In Sec. \ref{sec2}, we introduce the theoretical framework of $f(T, \mathcal{T})$ gravity and the basics of the FLRW model. Sec. \ref{sec3} discusses the parametrization of bounce cosmology within the context of $f(T, \mathcal{T})$ gravity. In Sec. \ref{sec4}, we analyze the energy conditions, examining their implications in gravitational theory. Finally, Sec. \ref{sec5} provides concluding remarks.

\section{Theoretical Background of $f(T, \mathcal{T})$ Gravity}\label{sec2}

The primary field in the considered gravity framework is the vierbein, denoted as $\mathbf{e}_A(x^\mu)$. At each point  $x^\mu$ in spacetime, the vierbein defines an orthonormal basis for the tangent space, adhering to the condition $\mathbf{e}_A \cdot \mathbf{e}_B = \eta_{AB}$, where $\eta_{AB} = \text{diag}(1, -1, -1, -1)$ is the Minkowski metric. The vierbein can also be expressed as a linear combination of the coordinate basis, specifically $\mathbf{e}_A = e^\mu_A \partial_\mu$. This allows for the following relationship:
\begin{equation}
g_{\mu\nu}(x)=\eta_{AB}\, e^A_\mu (x)\, e^B_\nu (x).
\end{equation}

In teleparallel gravitational theory, a key concept is teleparallelism, where the vierbein components at different points are parallel, giving the theory its name. In this framework, the Weitzenb\"{o}ck connection, represented by $\Gamma^\lambda_{\nu\mu}\equiv e^\lambda_A\:\partial_\mu e^A_\nu$, is used. This connection leads to zero curvature but a non-zero torsion, distinguishing it from the Levi-Civita connection, which results in zero torsion. The associated torsion tensor is defined as \cite{RR30}
\begin{equation}
\label{torsion2}
{T}^\lambda_{\:\mu\nu}=\Gamma^\lambda_{
\nu\mu}-%
\Gamma^\lambda_{\mu\nu}
=e^\lambda_A\:(\partial_\mu
e^A_\nu-\partial_\nu e^A_\mu).
\end{equation}

Furthermore, the contortion tensor and the corresponding superpotential tensor are defined as follows:
\begin{equation}
K^{\mu\nu}{}_{\rho}\equiv-\frac{1}{2}\Big(T^{\mu\nu}{}_{\rho}
-T^{\nu\mu}{}_{\rho}-T_{\rho}{}^{\mu\nu}\Big),    
\end{equation}
\begin{equation}
\label{s}
S_{\rho}{}^{\mu\nu}\equiv\frac{1}{2}\Big(K^{\mu\nu}{}_{\rho}
+\delta^\mu_\rho
\:T^{\alpha\nu}{}_{\alpha}-\delta^\nu_\rho\:
T^{\alpha\mu}{}_{\alpha}\Big).    
\end{equation}

Using Eqs. (\ref{torsion2}) and (\ref{s}), the torsion scalar $T$ is defined as \cite{Arcos/2004,Maluf/2013}
\begin{equation}
\label{torsionscalar}
T=S_{\rho}{}^{\mu\nu} T^{\rho}_{\ \mu \nu}=\frac{1}{4}
T^{\rho \mu \nu}
T_{\rho \mu \nu}
+\frac{1}{2}T^{\rho \mu \nu }T_{\nu \mu\rho}
-T_{\rho \mu}{}^{\rho }T^{\nu\mu}{}_{\nu}.
\end{equation}

Thus, when $T$ is used in the action and variations are taken with respect to the vierbeins, the resulting field equations coincide with those of GR. This correspondence gives the theory its name: the teleparallel equivalent of GR (TEGR). TEGR serves as a foundational framework for exploring various gravitational modifications. One such modification involves extending the action to $T + f(T)$, leading to $f(T)$ gravity. A further extension, known as $f(T, \mathcal{T})$ gravity, has an action given by \cite{Harko_fTT}
\begin{equation}
S= \frac{1}{16\,\pi\,G}\,\int d^{4}x\,e\,\left[T+f(T,\mathcal{T})\right]%
+\int d^{4}x\,e\,\mathcal{L}_{m}.
\label{action1}
\end{equation}%

Here, $e = \det(e^A_\mu) = \sqrt{-g}$ represents the determinant of the vierbein, $G$ is Newton's gravitational constant, $\mathcal{T}$ denotes the trace of the energy-momentum tensor $T^{\mu}_{\ \nu}$, and $\mathcal{L}_{m}$ is the matter Lagrangian density. The field equations resulting from varying the action (\ref{action1}) with respect to the vierbeins are given by
\begin{multline}
\left(1+f_{T}\right) \left[e^{-1} \partial_{\mu}{(e
e^{\alpha}_{A}
S_{\alpha}^{~\rho \mu})}-e^{\alpha}_{A} T^{\mu}_{~\nu \alpha} S_{\mu}^{~\nu
\rho}\right]+ 
\left(f_{TT} \partial_{\mu}{T}+f_{T\mathcal{T}} \partial_{\mu}{%
\mathcal{T}}\right) e^{\alpha}_{A} S_{\alpha}^{~\rho \mu}+ e_{A}^{\rho}
\left(\frac{f+T}{4}\right)-\\
f_{\mathcal{T}} \left(\frac{e^{\alpha}_{A} \mathcal{T}%
{}_{\alpha}^{~~\rho}+p e^{\rho}_{A}}{2}\right)=4\pi G e^{\alpha}_{A}
\mathcal{T}_{\alpha}{}^{\rho},
\label{geneoms}
\end{multline}
where $f_{\mathcal{T}}=\partial{f}/\partial{\mathcal{T}}$ and
$f_{T\mathcal{T%
}}=\partial^2{f}/\partial{T} \partial{\mathcal{T}}$.

Recent CMB data indicate that our universe is homogeneous—meaning it has the same properties at every point in space—and isotropic, implying it looks the same in all directions on large scales. Therefore, in the analysis presented here, we assume a flat Friedmann-Lemaître-Robertson-Walker (FLRW) background geometry in Cartesian coordinates, with the metric given by \cite{ryden/2003}
\begin{equation}
\label{FLRW}
ds^2 = dt^2 - a(t)^2 \left( dx^2 + dy^2 + dz^2 \right).
\end{equation}

In this context, $a(t)$ represents the scale factor of the universe, which quantifies the relative expansion of the universe over time. For this metric, the torsion scalar is $T = -6H^2$, where $H$ is the Hubble parameter. The vierbein field is taken in its diagonal form, given by
\begin{equation}
e_{\mu}^A = \text{diag}(1, a(t), a(t), a(t)),
\end{equation}
for the aforementioned metric. In addition, within the framework of a perfect fluid model—commonly used in cosmology to represent an ideal fluid under conditions where viscosity and heat conduction are absent—the energy-momentum tensor is expressed as:
\begin{equation}
\mathcal{T}_{\mu\nu} = (\rho + p)u_\mu u_\nu - p g_{\mu\nu},
\end{equation}
where $\rho$ is the energy density, $p$ is the pressure, and $u^\mu=(1,0,0,0)$ represents the four-velocity of the fluid elements. The trace of the stress-energy tensor, denoted as $\mathcal{T}$, is given by $\mathcal{T} = \rho - 3p$.

The Friedmann equations describe the evolution of the universe's scale factor, relating its expansion to the energy content and curvature of spacetime. The corresponding modified Friedmann equations for the metric (\ref{FLRW}) and an arbitrary $f(T, \mathcal{T})$ function are given by \cite{Harko_fTT}
\begin{equation}\label{F1}
H^2 =\frac{8\pi G}{3}\rho - \frac{1}{6}\left(f+12H^2f_T \right)+f_\mathcal{T}\left(\frac{\rho+p}{3} \right),
\end{equation}

\begin{equation}
\label{F2}
\dot{H}= -4\pi G(\rho+p)-\dot{H}(f_T-12H^2 f_{TT})-H(\dot{\rho}-3\dot{p}) f_{T \mathcal{T }} - f_\mathcal{T}\left(\frac{\rho+p}{2} \right),
\end{equation}
where the dot symbol ($\cdot$) denotes a time derivative.

\section{Cosmological $f(T, \mathcal{T})$ Models with a Bouncing Scenario}\label{sec3}

For a successful bouncing scenario in cosmology, several key conditions must be met to ensure that the universe undergoes a smooth transition from a contracting phase to an expanding phase without encountering singularities. In the context of \( f(T, \mathcal{T}) \) gravity, the following conditions are generally required for a successful bounce \cite{Cai/2007}:

1. \textbf{Scale factor evolution}: In the contracting phase (before the bounce), the scale factor $a(t)$ must decrease ($\dot{a} < 0$), while in the expanding phase (after the bounce), it must increase ($ \dot{a} > 0 $). The scale factor must be finite and non-vanishing at the bouncing epoch, thus resolving the initial singularity problem. Specifically, at the bounce, $ \dot{a} = 0 $.

2. \textbf{Hubble parameter behavior}: The Hubble parameter $H(t)$ must transition from negative values before the bounce ($H(t) < 0$) to positive values after the bounce ($H(t) > 0$), with $H = 0$ exactly at the bounce point. For a successful bounce scenario, a rapid violation of the Null Energy Condition (NEC) must occur in the vicinity of the bounce. To achieve this, $\dot{H} = -4\pi G(1+\omega)\rho > 0 $ in the bouncing region, which implies that the EoS parameter $\omega$ must satisfy $\omega < -1 $ near the bounce.

3. \textbf{EoS Behavior}: The EoS parameter $\omega$ must cross the phantom divide line $\omega = -1$ during the bounce, transitioning from quintessence-like behavior (with $\omega > -1$) to phantom-like behavior (with $\omega < -1$) and back.

Together, these conditions ensure a successful bouncing scenario in cosmological models, resolving the issue of singularities and allowing for a smooth transition between the contraction and expansion phases. Motivated by these conditions, we assume the following parameterization for the scale factor of bounce cosmology \cite{Zubair/2023,Shabani/2018}:
\begin{equation}
\label{at}
a(t)=\sqrt{a_0^2+\gamma ^2 t^2},    
\end{equation}
where $a_0$ is a constant representing the initial scale factor at $t = 0$, and $\gamma$ is a positive parameter that controls the rate of expansion and contraction. This form of the scale factor ensures a smooth transition through the bounce, with the scale factor being finite and non-vanishing at the bouncing epoch, and the Hubble parameter $H(t)$ crossing from negative to positive values as required by the bounce conditions. The model also satisfies the necessary criteria to resolve the initial singularity problem, as the scale factor grows smoothly from the contracting phase to the expanding phase, avoiding any infinite curvature or energy densities.

The corresponding Hubble parameter associated with the scale factor (\ref{at}) is given by
\begin{equation}
H(t)=\frac{\dot{a}(t)}{a(t)}=\frac{\gamma ^2 t}{a_0^2+\gamma ^2 t^2}.   \label{Ht} 
\end{equation}

The deceleration parameter is expressed as
\begin{equation}
q(t)=-1+\frac{d}{dt}\left( \frac{1}{H\left( t\right) }\right)=-\frac{a_0^2}{\gamma ^2 t^2}   
\end{equation}

\begin{figure}[h]
     \centering
     \begin{subfigure}[b]{0.33\textwidth}
         \centering
         \includegraphics[width=\textwidth]{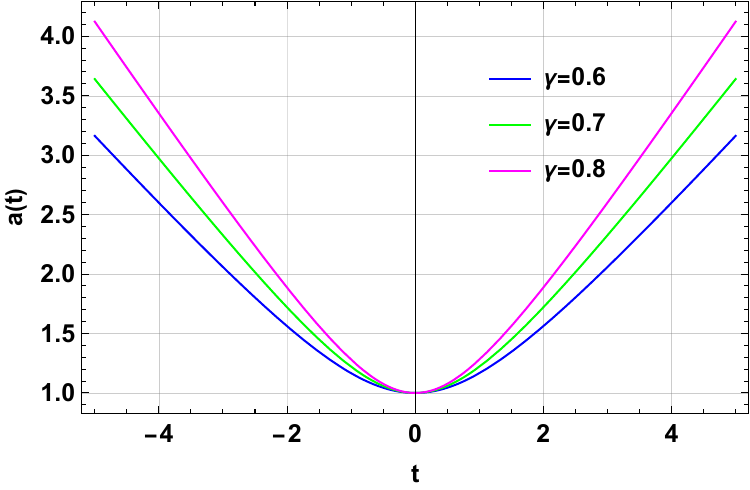}
         \caption{Scale factor}
         \label{F_a}
     \end{subfigure}
     \hfill
     \begin{subfigure}[b]{0.33\textwidth}
         \centering
         \includegraphics[width=\textwidth]{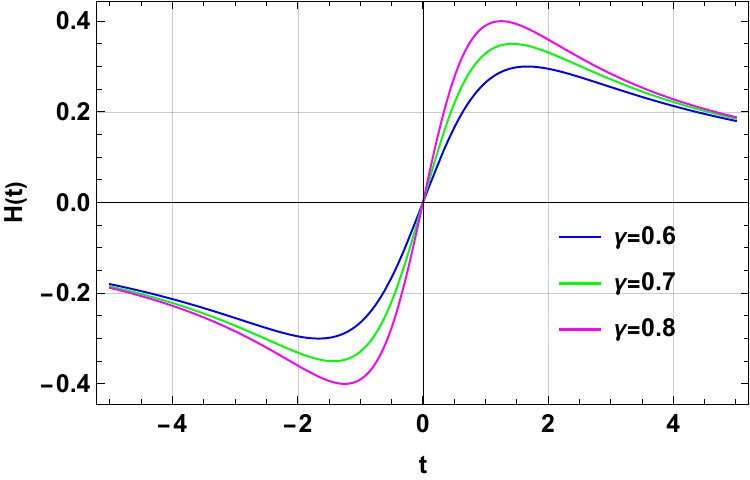}
         \caption{Hubble parameter}
         \label{F_H}
     \end{subfigure}
     \hfill
     \begin{subfigure}[b]{0.33\textwidth}
         \centering
         \includegraphics[width=\textwidth]{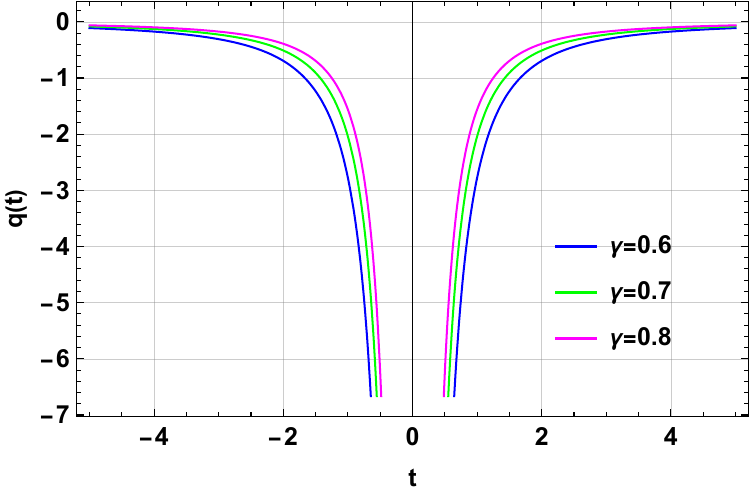}
         \caption{Deceleration parameter}
         \label{F_q}
     \end{subfigure}
        \caption{The evolution of the scale factor, Hubble parameter, and deceleration parameter as functions of cosmic time for $a_0 = 1$, evaluated at various values of $\gamma$.}
\end{figure}

Fig. \ref{F_a} shows the behavior of the scale factor as a function of cosmic time for different values of the bouncing parameter $\gamma = 0.6, 0.7, 0.8$. As $\gamma$ increases, the slope of the scale factor becomes steeper, indicating a faster rate of expansion on both sides of the bounce. This symmetric expansion around the bounce point $t = 0$ illustrates the initial contraction, a smooth transition through the bounce, and subsequent expansion, with $\gamma$ significantly influencing the dynamics of the scale factor curve. Fig. \ref{F_H} illustrates the corresponding behavior of the Hubble parameter in the bounce scenario. The Hubble parameter fulfills the requirements for a bouncing cosmology by starting with negative values, crossing through zero at $t = 0$, and then becoming positive. This transition is essential for describing the bounce and subsequent expansion. Fig. \ref{F_q} presents the evolution of the deceleration parameter. The negative range of the deceleration parameter signals an accelerated phase of expansion. Near the bounce, the deceleration parameter exhibits a singularity, which is a characteristic feature of the selected bounce model. With these favorable behaviors of the key cosmological parameters, we can use the chosen parameterization to further evaluate the cosmological $f(T,\mathcal{T})$ models.

\subsection{Model I: $f(T,\mathcal{T})=\alpha T + \beta \mathcal{T}$}

The first form, $f(T, \mathcal{T}) = \alpha T + \beta \mathcal{T}$ (where $\alpha$ and $\beta$ are constants), is a linear model motivated by simplicity and compatibility with both GR and observational constraints. Given that modifications to GR must be subtle to remain consistent with solar system tests and cosmological data, this linear model closely approximates a linear $f(T)$ function, which is effectively a TEGR with an added term depending on $\mathcal{T}$ \cite{fTT1,fTT2,fTT3}.

Thus, we obtain $f_{T} = \alpha$, $f_{\mathcal{T}} = \beta$, $f_{TT} = 0$, and $f_{T \mathcal{T}} = 0$. Using Eqs. (\ref{F1}) and (\ref{F2}), we then derive the expressions for the energy density, pressure, and EoS parameter ($\omega = \frac{p}{\rho}$) as follows:
\begin{eqnarray} \label{rho1}
\rho&=&-\frac{3 (\alpha +1) \left((2 \beta +3) H^2+5 \beta  \dot{H}\right)}{4 \beta  (\beta +1)-3}, \\
\label{p1}
p&=&\frac{3 (\alpha +1) \left((2 \beta +3) H^2+(\beta +2) \dot{H}\right)}{4 \beta  (\beta +1)-3}, \\
\omega&=&-\frac{(2 \beta +3) H^2+(\beta +2) \dot{H}}{(2 \beta +3) H^2+5 \beta  \dot{H}}.
\end{eqnarray}

Now, based on the parameterization of the scale factor in bounce cosmology, we obtain
\begin{eqnarray} 
\rho&=&\frac{3 (\alpha +1) \gamma ^2 \left(3 (\beta -1) \gamma ^2 t^2-5 \beta \right)}{(4 \beta  (\beta +1)-3) \left(\gamma ^2 t^2+1\right)^2}, \\
p&=&\frac{3 (\alpha +1) \gamma ^2 \left(\beta +(\beta +1) \gamma ^2 t^2+2\right)}{(4 \beta  (\beta +1)-3) \left(\gamma ^2 t^2+1\right)^2}, \\
\omega&=&\frac{\beta +(\beta +1) \gamma ^2 t^2+2}{3 (\beta -1) \gamma ^2 t^2-5 \beta }.
\end{eqnarray}

\begin{figure}[h]
     \centering
     \begin{subfigure}[b]{0.33\textwidth}
         \centering
         \includegraphics[width=\textwidth]{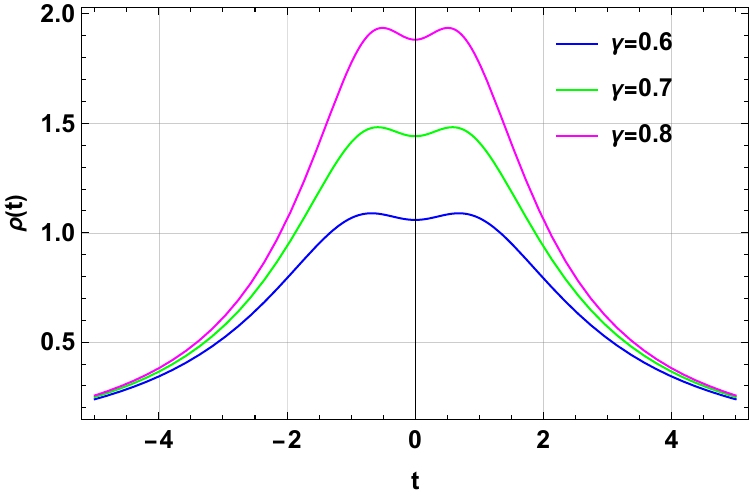}
         \caption{Energy density}
         \label{F_rho1}
     \end{subfigure}
     \hfill
     \begin{subfigure}[b]{0.33\textwidth}
         \centering
         \includegraphics[width=\textwidth]{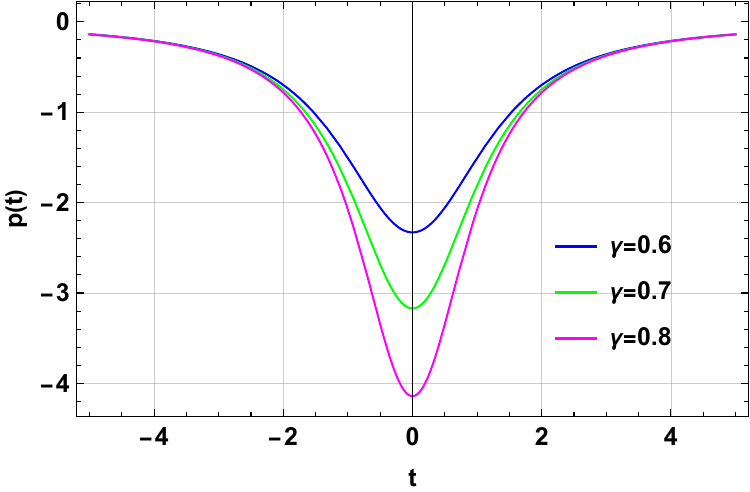}
         \caption{Pressure}
         \label{F_p1}
     \end{subfigure}
     \hfill
     \begin{subfigure}[b]{0.33\textwidth}
         \centering
         \includegraphics[width=\textwidth]{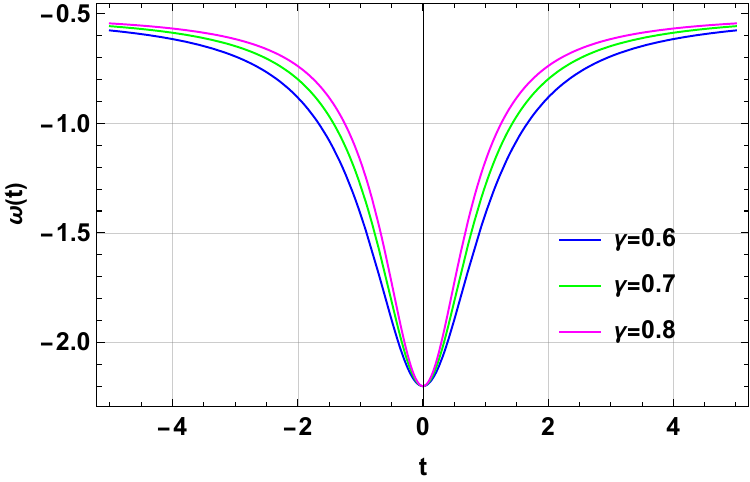}
         \caption{EoS parameter}
         \label{F_w1}
     \end{subfigure}
        \caption{The evolution of the energy density, pressure, and EoS parameter as functions of cosmic time for $a_0 =\alpha = 1$, and $\beta = 0.2$, evaluated at different values of $\gamma$ (Model I).}
        \label{F_M1}
\end{figure}

In Fig. \ref{F_rho1}, the energy density shows a dramatic increase as the universe approaches the bounce point, reaching a peak before slightly decreasing at the bounce epoch. After the bounce, the energy density rises slightly again but then gradually decreases as cosmic time progresses. This behavior indicates the high energy fluctuations during the bounce and their subsequent relaxation post-bounce \cite{K_B1,K_B2,K_B3}. In Fig. \ref{F_p1}, the pressure is negative throughout the cosmic evolution \cite{Agrawal/2021}. For all values of the bounce parameter $\gamma$, the pressure starts off with a small negative value in the pre-bounce phase, increases to a more significant negative value at the bounce, and then decreases again to a smaller negative value in the post-bounce era. This indicates that the universe undergoes a contracting phase with negative pressure, which is characteristic of a phase leading to the bounce, followed by a brief period of low pressure after the bounce. In Fig. \ref{F_w1}, the EoS parameter for different values of $\gamma$ demonstrates a significant sensitivity to the bounce parameter. For non-zero values of $a_0$, the EoS parameter shows negative values for all $\gamma$ values. Specifically, the EoS parameter corresponding to Model I crosses the phantom divide line ($\omega = -1$) near the bounce and enters the phantom region, while it remains in the quintessence phase away from the bounce. Both the pre- and post-bounce regions display similar patterns, indicating the transition from a decelerating to an accelerating phase during the bounce.

\subsection{Model II: $f(T,\mathcal{T})=\alpha  \sqrt{-T} + \beta \mathcal{T}$}

The second form, $f(T, \mathcal{T}) = \alpha \sqrt{-T} + \beta \mathcal{T}$, is a non-linear model that introduces a square-root dependence on $T$. This model is motivated by the need for a richer structure that can capture effects potentially leading to cosmic acceleration, even without a cosmological constant, and offers more flexibility to address observational data \cite{fTT4}.

As a result, we find $f_{T} = -\frac{\alpha }{2 \sqrt{-T}}$, $f_{\mathcal{T}} = \beta$, $f_{TT} =-\frac{\alpha }{4 (-T)^{3/2}}$, and $f_{T \mathcal{T}} = 0$. Applying Eqs. (\ref{F1}) and (\ref{F2}), we then derive the expressions for the energy density, pressure, and the EoS parameter as follows:
\begin{eqnarray} \label{rho2}
\rho&=&-\frac{3 \left((2 \beta +3) H^2+5 \beta  \dot{H}\right)}{4 \beta  (\beta +1)-3}, \\
\label{p2}
p&=&\frac{3 (2 \beta +3) H^2+3 (\beta +2) \dot{H}}{4 \beta  (\beta +1)-3}, \\
\omega&=&-\frac{(2 \beta +3) H^2+(\beta +2) \dot{H}}{(2 \beta +3) H^2+5 \beta  \dot{H}}.
\end{eqnarray}

In addition, based on the parameterization of the scale factor in bounce cosmology, we obtain
\begin{eqnarray} 
\rho&=&\frac{9 (\beta -1) \gamma ^4 t^2-15 \beta  \gamma ^2}{(4 \beta  (\beta +1)-3) \left(\gamma ^2 t^2+1\right)^2}, \\
p&=&\frac{3 \gamma ^2 \left(\beta +(\beta +1) \gamma ^2 t^2+2\right)}{(4 \beta  (\beta +1)-3) \left(\gamma ^2 t^2+1\right)^2}, \\
\omega&=&\frac{\beta +(\beta +1) \gamma ^2 t^2+2}{3 (\beta -1) \gamma ^2 t^2-5 \beta }.
\end{eqnarray}

\begin{figure}[h]
     \centering
     \begin{subfigure}[b]{0.33\textwidth}
         \centering
         \includegraphics[width=\textwidth]{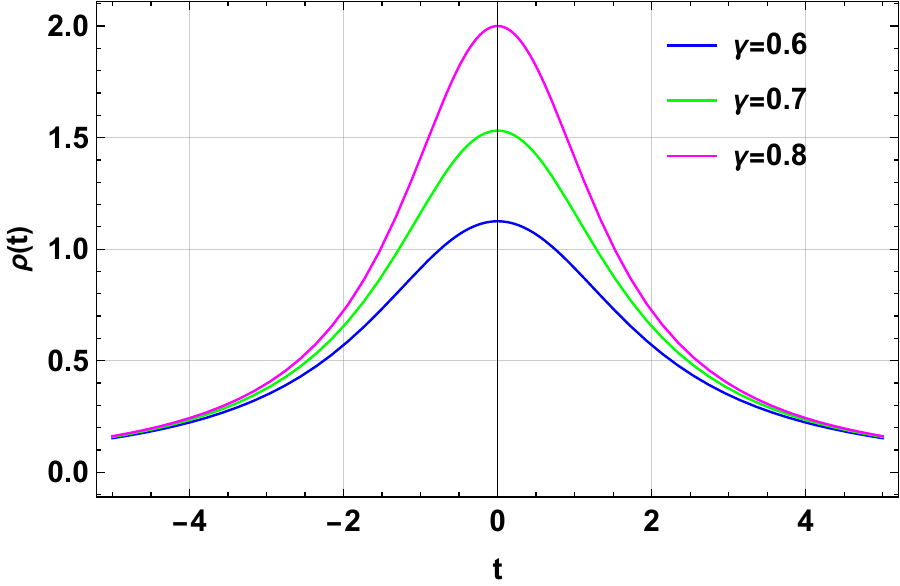}
         \caption{Energy density}
         \label{F_rho2}
     \end{subfigure}
     \hfill
     \begin{subfigure}[b]{0.33\textwidth}
         \centering
         \includegraphics[width=\textwidth]{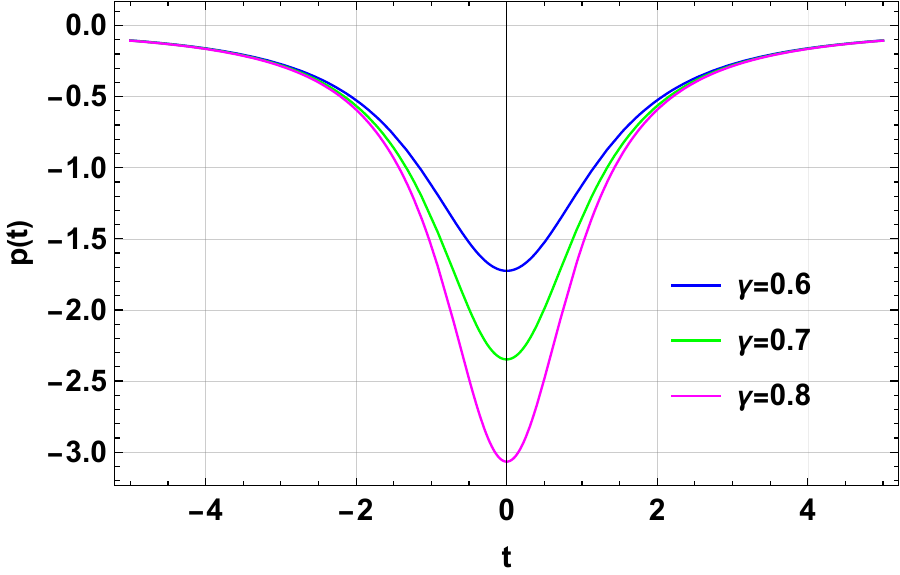}
         \caption{Pressure}
         \label{F_p2}
     \end{subfigure}
     \hfill
     \begin{subfigure}[b]{0.33\textwidth}
         \centering
         \includegraphics[width=\textwidth]{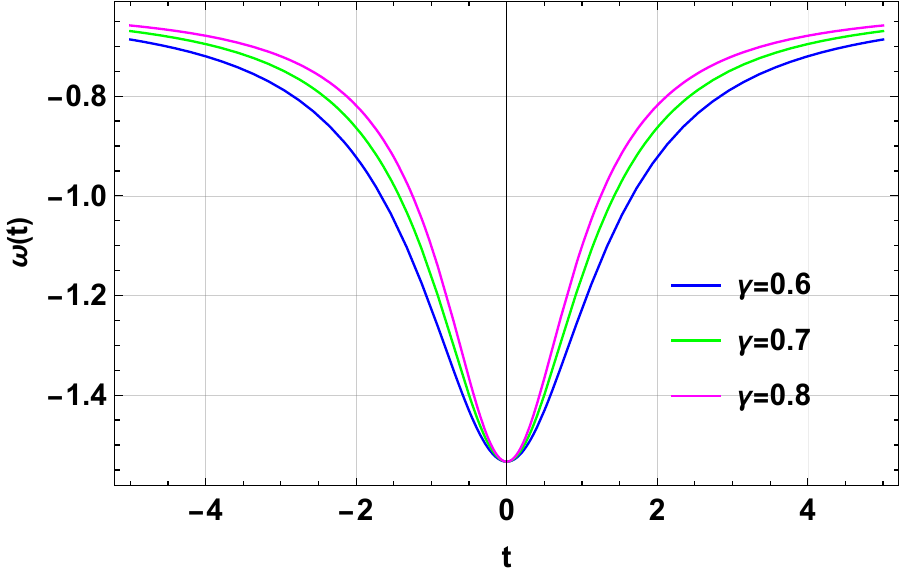}
         \caption{EoS parameter}
         \label{F_w2}
     \end{subfigure}
        \caption{The evolution of the energy density, pressure, and EoS parameter as functions of cosmic time for $a_0 =\alpha = 1$, and $\beta = 0.3$, evaluated at different values of $\gamma$ (Model II).}
        \label{F_M2}
\end{figure}

In Fig. \ref{F_rho2}, the energy density increases significantly as the universe approaches the bounce, peaking sharply at the bounce epoch. After the bounce, it gradually decreases with cosmic time. This pattern shows that, as in Model I, the energy density experiences a strong fluctuation at the bounce, though it stabilizes afterward. In Fig. \ref{F_p1}, the pressure remains negative throughout the evolution. For all values of the bounce parameter $\gamma$, the pressure begins with a slightly negative value in the pre-bounce phase, increases to a larger negative value at the bounce, but remains lower compared to Model I. Post-bounce, the pressure declines back to a small negative value. This behavior suggests a consistent negative pressure that is essential for driving the bounce, but with smaller fluctuations than in Model I. In Fig. \ref{F_w1}, the EoS parameter is negative across all values of $\gamma$. Like Model I, the EoS parameter for Model II crosses the phantom divide line ($\omega = -1$) near the bounce, entering the phantom region; however, it reaches a smaller negative value than in Model I. This crossing indicates a temporary transition into the phantom regime, followed by a return to less extreme values, highlighting a softer bounce effect in Model II compared to Model I.

\section{Energy conditions}
\label{sec4}

Energy conditions are a set of constraints on the relationship between energy density and pressure to ensure the physically meaningful behavior of spacetime. These conditions imply that energy density cannot be negative and that gravity maintains its attractive nature. They assert that specific linear combinations of pressure and density cannot be negative, which is crucial in studies of cosmology, wormholes, and black hole thermodynamics \cite{ECs1,ECs2,ECs3,ECs4,ECs5}. These conditions are derived from Raychaudhuri's equation. The four main energy conditions are as follows:

\begin{itemize}

\item \textbf{Null energy condition (NEC):}
\begin{equation}
  \mathcal{T}_{\mu\nu} k^\mu k^\nu \geq 0 \quad \text{for all null vectors } k^\mu \Longleftrightarrow \rho + p \geq 0.
\end{equation}
  The NEC states that for any lightlike (null) path in spacetime, the projection of the energy-momentum tensor along that path must be non-negative. This condition is often crucial for the stability of spacetime and is frequently applied in studies of cosmological evolution and black hole physics.

\item \textbf{Weak energy condition (WEC):}
\begin{equation}
  \mathcal{T}_{\mu\nu} v^\mu v^\nu \geq 0 \quad \text{for all timelike vectors } v^\mu \Longleftrightarrow \rho \geq 0, \; \rho + p \geq 0.
\end{equation}
  The WEC ensures that any observer moving along a timelike path will measure a non-negative energy density \cite{ECs3}. This condition implies that matter possesses positive energy, which is essential for physically realistic matter configurations.

\item \textbf{Dominant energy condition (DEC):}
\begin{equation}
  \mathcal{T}_{\mu\nu} v^\mu v^\nu \geq 0 \quad \text{and} \quad \mathcal{T}_{\mu\nu} v^\nu \text{ is causal for all timelike vectors } v^\mu \Longleftrightarrow \rho \geq 0, \; |p| \leq \rho.
\end{equation}
  The DEC requires that energy density be non-negative and that the energy flow vector $T_{\mu\nu} v^\nu$ is causal (i.e., it does not exceed the speed of light). This condition maintains causality and is often used to exclude unphysical matter forms with superluminal properties.

\item \textbf{Strong energy condition (SEC):}
\begin{equation}
  \mathcal{T}_{\mu\nu} v^\mu v^\nu - \frac{1}{2} \mathcal{T} g_{\mu\nu} v^\mu v^\nu \geq 0 \quad \text{for all timelike vectors } v^\mu \Longleftrightarrow \rho + 3p \geq 0.
\end{equation}
  The SEC implies that gravitational effects are attractive, requiring that the combination of energy density and pressure ($\rho + 3p $) be non-negative. This condition is typically used in cosmology and gravitational collapse scenarios to study the deceleration of expansion in the universe \cite{ECs3}.

  \end{itemize}

These conditions are instrumental in assessing the viability of cosmological models, including bouncing cosmologies, as they help determine the nature of energy density and pressure and can highlight non-standard behaviors necessary for phenomena like cosmic bounces.

\subsection{Model I: $f(T,\mathcal{T})=\alpha T + \beta \mathcal{T}$}

Using Eqs. (\ref{rho1}) and (\ref{p1}), we derive the following expressions for the energy conditions for the $f(T,\mathcal{T})$ Model I:
\begin{eqnarray} 
\rho+p&=&\frac{6 (\alpha +1) \gamma ^2 \left(\gamma ^2 t^2-1\right)}{(2 \beta +3) \left(\gamma ^2 t^2+1\right)^2} \geq 0, \\
\rho-p&=&\frac{6 (\alpha +1) \gamma ^2 \left(-3 \beta +(\beta -2) \gamma ^2 t^2-1\right)}{(4 \beta  (\beta +1)-3) \left(\gamma ^2 t^2+1\right)^2} \geq 0, \\
\rho+3p&=&\frac{6 (\alpha +1) \gamma ^2 \left(\beta  \left(3 \gamma ^2 t^2-1\right)+3\right)}{(4 \beta  (\beta +1)-3) \left(\gamma ^2 t^2+1\right)^2} \geq 0.
\end{eqnarray}

\begin{figure}[h]
     \centering
     \begin{subfigure}[b]{0.33\textwidth}
         \centering
         \includegraphics[width=\textwidth]{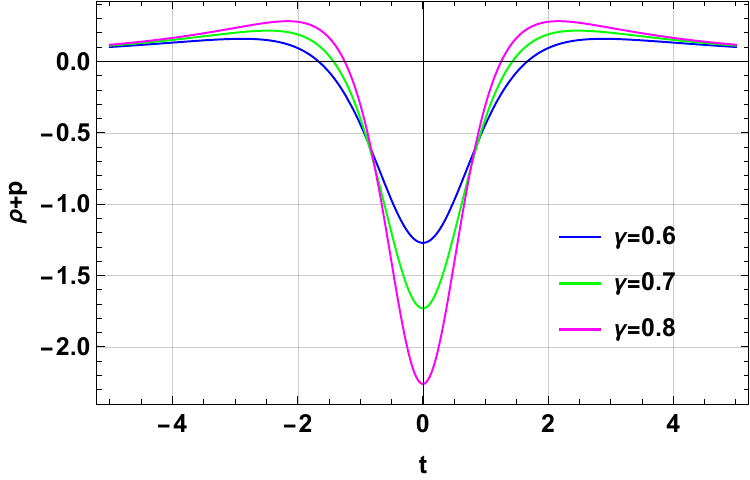}
         \caption{NEC}
         \label{F_NEC1}
     \end{subfigure}
     \hfill
     \begin{subfigure}[b]{0.33\textwidth}
         \centering
         \includegraphics[width=\textwidth]{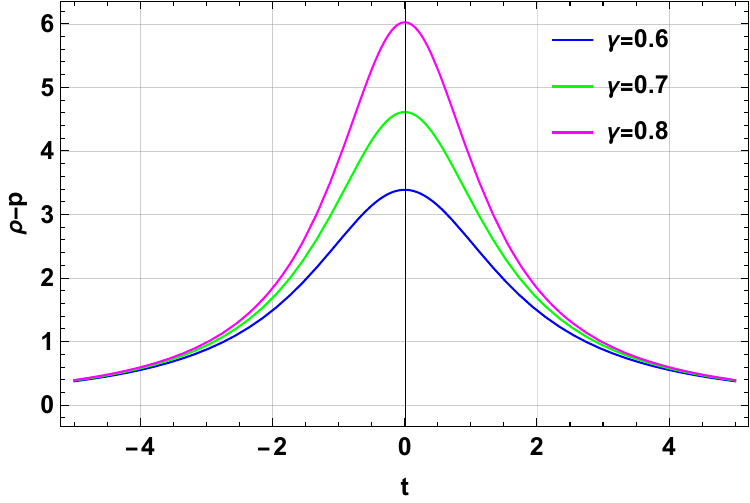}
         \caption{DEC}
         \label{F_DEC1}
     \end{subfigure}
     \hfill
     \begin{subfigure}[b]{0.33\textwidth}
         \centering
         \includegraphics[width=\textwidth]{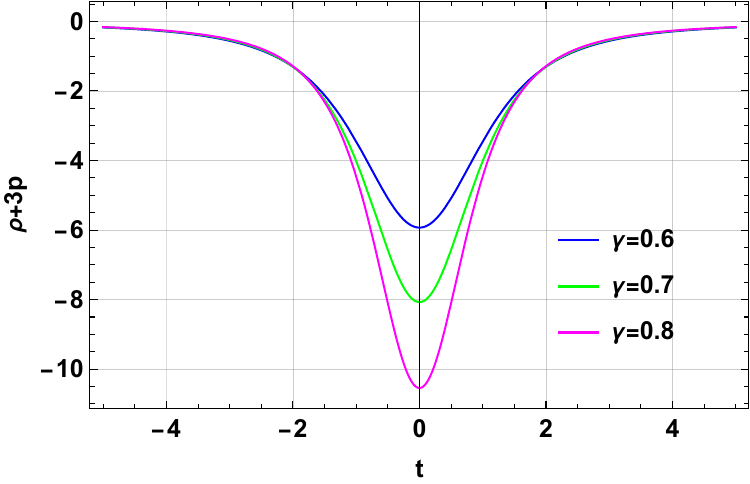}
         \caption{SEC}
         \label{F_SEC1}
     \end{subfigure}
        \caption{The evolution of the energy conditions as functions of cosmic time for $a_0 =\alpha = 1$, and $\beta = 0.2$, evaluated at different values of $\gamma$ (Model I).}
\end{figure}

For a successful bounce, the NEC must be violated, as shown in Fig. \ref{F_NEC1} \cite{Cai/2007}. This violation is essential because it allows the EoS parameter $\omega$ to enter the phantom regime ($ \omega < -1 $) near the bounce, facilitating a reversal of contraction to expansion. The NEC violation thus provides the key to enabling the bounce and the avoidance of singularity. The graph highlights a distinct NEC violation around the bounce epoch, supporting the non-standard expansion phase of the universe. The SEC is also violated in this model, as depicted in Fig. \ref{F_SEC1}, which follows directly from the NEC violation. This SEC violation contributes to the model's evolution within the phantom phase, where $\omega < -1$, indicating that the gravitational effects are repulsive rather than attractive. This violation is consistent with the accelerated expansion of the universe seen around the bounce \cite{ECs4}. Furthermore, Fig. \ref{F_DEC1} shows that the DEC is satisfied throughout the cosmic evolution for the range of parameter $\gamma$. This satisfaction implies that the energy density remains positive and causality is preserved, which aligns with the behavior of a perfect fluid-like matter distribution. The DEC’s positive behavior suggests a realistic matter configuration without superluminal energy propagation. 

\subsection{Model II: $f(T,\mathcal{T})=\alpha  \sqrt{-T} + \beta \mathcal{T}$}

From Eqs. (\ref{rho2}) and (\ref{p2}), we obtain the following expressions for the energy conditions in the $f(T,\mathcal{T})$ Model II:
\begin{eqnarray} 
\rho+p&=&\frac{6 (\alpha +1) \gamma ^2 \left(\gamma ^2 t^2-1\right)}{(2 \beta +3) \left(\gamma ^2 t^2+1\right)^2} \geq 0, \\
\rho-p&=&\frac{6 (\alpha +1) \gamma ^2 \left(-3 \beta +(\beta -2) \gamma ^2 t^2-1\right)}{(4 \beta  (\beta +1)-3) \left(\gamma ^2 t^2+1\right)^2} \geq 0, \\
\rho+3p&=&\frac{6 (\alpha +1) \gamma ^2 \left(\beta  \left(3 \gamma ^2 t^2-1\right)+3\right)}{(4 \beta  (\beta +1)-3) \left(\gamma ^2 t^2+1\right)^2} \geq 0.
\end{eqnarray}

\begin{figure}[h]
     \centering
     \begin{subfigure}[b]{0.33\textwidth}
         \centering
         \includegraphics[width=\textwidth]{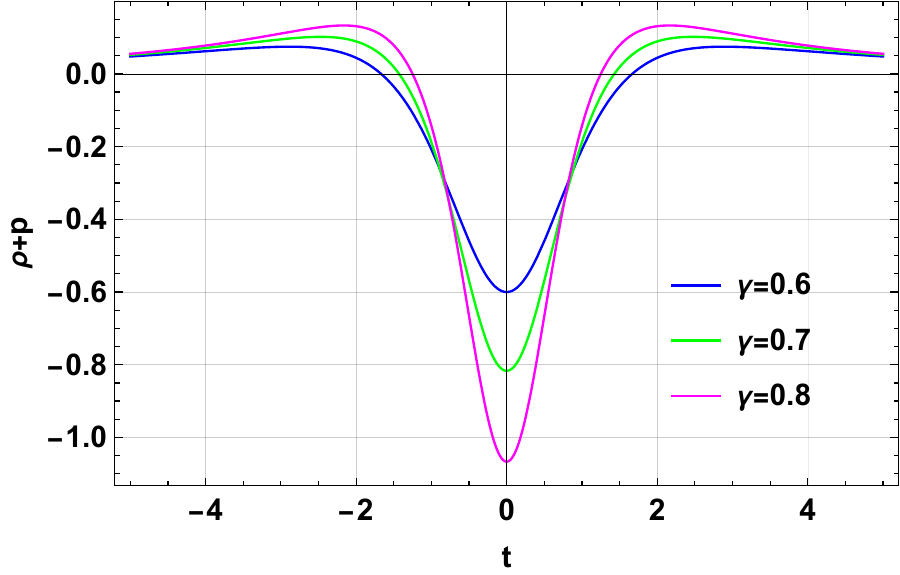}
         \caption{NEC}
         \label{F_NEC2}
     \end{subfigure}
     \hfill
     \begin{subfigure}[b]{0.33\textwidth}
         \centering
         \includegraphics[width=\textwidth]{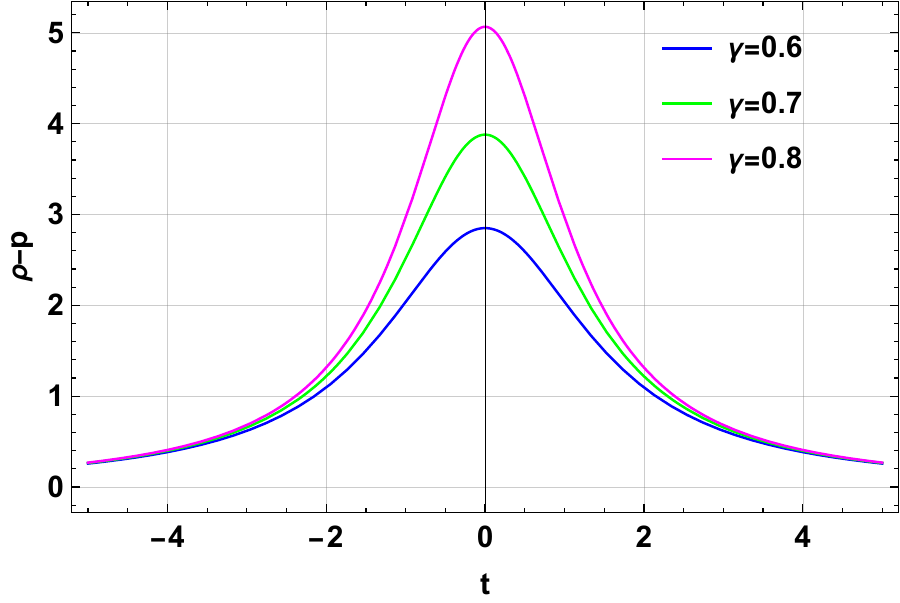}
         \caption{DEC}
         \label{F_DEC2}
     \end{subfigure}
     \hfill
     \begin{subfigure}[b]{0.33\textwidth}
         \centering
         \includegraphics[width=\textwidth]{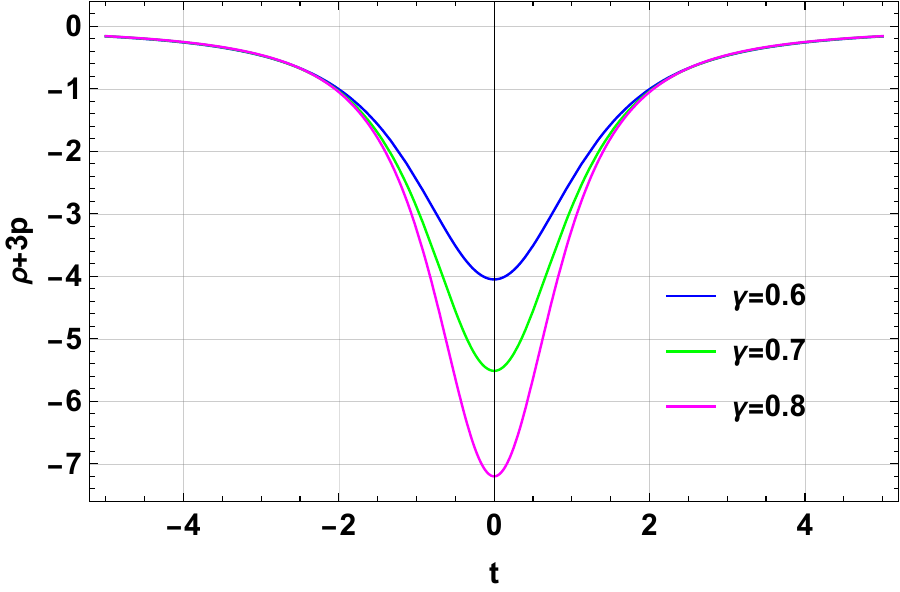}
         \caption{SEC}
         \label{F_SEC2}
     \end{subfigure}
        \caption{The evolution of the energy conditions as functions of cosmic time for $a_0 =\alpha = 1$, and $\beta = 0.3$, evaluated at different values of $\gamma$ (Model II).}
\end{figure}

For Model II (non-linear), we observe similar results to Model I regarding the energy conditions: The NEC is violated near the bounce as the EoS parameter enters the phantom region ($ \omega < -1 $), supporting the occurrence of a non-singular bounce, as in Model I (Fig. \ref{F_NEC2}). Also, the SEC is violated, following the NEC violation, which allows the model to enter the phantom phase and results in an accelerated expansion, mirroring Model I (Fig. \ref{F_SEC2}). The DEC remains satisfied, indicating a stable, perfect fluid-like matter distribution, despite the NEC and SEC violations, as seen in Model I (Fig. \ref{F_SEC2}). Therefore, as with Model I, there is a uniform shift in energy conditions around the bounce, ensuring non-singular behavior and a smooth transition through the bounce epoch.

\section{Conclusion}\label{sec5}

The singularity and inflationary problems have recently presented substantial challenges for cosmologists aiming to comprehend the universe's origin and evolution. Due to limited observational data, researchers have explored bouncing cosmology as a viable alternative to traditional models, offering solutions to the singularity problem encountered in standard cosmological frameworks. Bouncing cosmology provides a different perspective on the early universe, enabling a non-singular cosmic evolution through a "bounce" rather than a singular beginning. In this study, we examined the feasibility of reproducing specific bouncing models within the $f(T, \mathcal{T})$ gravity framework, which allows for a broad coupling between the torsion scalar $T$ and the trace of the energy-momentum $\mathcal{T}$. Specifically, we considered two $f(T, \mathcal{T})$ models: a linear model given by $f(T, \mathcal{T}) = \alpha T + \beta \mathcal{T}$ \cite{fTT1,fTT2,fTT3} and a non-linear model $f(T, \mathcal{T}) = \alpha \sqrt{-T} + \beta \mathcal{T}$ \cite{fTT4}, where $\alpha$ and $\beta$ are arbitrary constants. Furthermore, the parameterization of the scale factor, $a(t) = \sqrt{a_0^2 + \gamma^2 t^2} $ \cite{Zubair/2023,Shabani/2018}, was employed to describe the bouncing behavior. By analyzing the dynamical behavior of parameters, we observed a cosmic bounce at $t = 0$, confirmed by the Hubble parameter reaching zero, which indicates the transition from contraction to expansion. One critical condition for achieving a bounce is the violation of the NEC, which we verified in the vicinity of the bounce. This NEC violation is essential for the EoS parameter to enter the phantom region ($\omega < -1$) near the bounce, allowing the model to undergo a smooth, non-singular bounce.

In both models, we found that the energy density exhibits a dramatic increase as the universe approaches the bounce, peaking at the bounce epoch and then decreasing as cosmic time progresses. Similarly, the pressure remains negative throughout the cosmic evolution, with a small negative value in the pre-bounce phase, a significant negative value at the bounce, and then a slight increase in the post-bounce era. The EoS parameter remains negative for all values of $\gamma$, and both models cross the phantom divide line ($\omega = -1$) near the bounce, entering the phantom region. This behavior is consistent for both positive and negative time zones. Regarding energy conditions, we observed that the NEC is violated near the bouncing epoch, which is a necessary condition for the realization of a non-singular bounce. This violation of the NEC leads to the violation of the SEC, forcing the model to evolve in the phantom phase ($\omega < -1$). The DEC remains satisfied in both models, indicating a consistent matter distribution throughout the cosmic evolution. In conclusion, both the linear and non-linear models of $f(T, \mathcal{T})$ gravity successfully reproduce the key features of bouncing cosmology, including the violation of energy conditions necessary for a non-singular bounce. These models offer a promising framework for addressing the singularity problem and the inflationary challenge in cosmology. Compared to matter bounce cosmologies in LQC \cite{Bojowald/2001} and $f(R)$ \cite{Bamba/2014} frameworks, the bounce obtained in $f(T,\mathcal{T})$ gravity is softer in the non-linear model and exhibits controlled violation of the NEC, similar in spirit to $f(T)$-based matter bounces \cite{B_T1}. Unlike $f(R)$, where higher-order derivatives may induce instabilities, our torsion-based approach avoids such issues while maintaining bounce viability. In this work, we focused on background cosmology. A full analysis of cosmological perturbations (as discussed in \cite{Cai/2010}), including scalar and tensor power spectra predictions, is left for future work. Such an extension will help assess compatibility with CMB data and distinguish $f(T,\mathcal{T})$ bouncing models from inflationary scenarios.

\section*{Data availability} 
This article does not include any new associated data.

%%%%%%%%%%%%%%%%%%%%%%%%%%%%%%%%%%%%%%%%%%%%%%%%%%%%%%%%%%%%%%%%%%%%%%%%%%%%%%%

\end{document}